\documentclass[preprint,secnumarabic,tightenlines,amssymb, amsmath,nobibnotes,
aps,nofootinbib,showpacs,]{revtex4-1}%

\usepackage{graphicx}
\usepackage{dcolumn}
\usepackage{bm}
\usepackage{caption}
\usepackage{amssymb}
\usepackage{setspace}
\usepackage{amsmath}
\usepackage{amsfonts}
\usepackage{verbatim}
\usepackage{MnSymbol}
\usepackage{hyperref}
\usepackage{multirow}
\usepackage{color}%
\definecolor{darkgreen}{rgb}{0,0.35,0}
\newcommand{\rc}{\textcolor{red}}

\definecolor{Rood}{rgb}{1, 0, 0}

\newcommand{\de}{\delta E}

\newcommand{\be}{\begin{equation}}
\newcommand{\ee}{\end{equation}}
\newcommand{\beq}{\begin{eqnarray}}
\newcommand{\eeq}{\end{eqnarray}}
\newcommand{\beqs}{\begin{eqnarray*}}
\newcommand{\eeqs}{\end{eqnarray*}}

\begin{document}

\title{Casimir energy for two and three superconducting \\  coupled cavities
}

\author{L. Rosa$^{1,2}$, S. Avino$^{2,3}$, E. Calloni$^{1,2}$, S. Caprara$^{ 4,5}$, M. De Laurentis$^1$, 
R. De Rosa$^{1,2}$, Giampiero Esposito$^2$, M. Grilli$^{4,5}$, E. Majorana$^6$,
G. P. Pepe$^7$, S. Petrarca$^{4,6}$, P. Puppo$^6$, P. Rapagnani$^{4,6}$, F. Ricci$^{4,6}$, 
C. Rovelli$^8$, P. Ruggi$^9$, N. L. Saini$^4$, C. Stornaiolo$^2$, F. Tafuri$^{1}$}

\affiliation
{$^1$  Universit\`a di Napoli Federico II, 
Dipartimento di Fisica ``Ettore Pancini'', 
Complesso Universitario di Monte S. Angelo,
Via Cintia Edificio 6, 80126 Napoli, Italy }
\affiliation
{$^2$ INFN Sezione di Napoli, 
Complesso Universitario di Monte S. Angelo,
Via Cintia Edificio 6, 80126 Napoli, Italy}
\affiliation
{$^3$  INO-CNR, 
Comprensorio Olivetti,\\
Via Campi Flegrei 34-80078 Pozzuoli (NA), Italy }
\affiliation
{$^4$  Universit\`a di Roma ``La Sapienza'', P.le A. Moro 2, I-00185, Roma, Italy}
\affiliation
{$^{5}$ISC-CNR and Consorzio Nazionale Interuniversitario per\\ le Scienze Fisiche della Materia (CNISM), Unit\`a di Roma ``La Sapienza'', P.le A. Moro 2, I-00185, Roma, Italy}
\affiliation
{$^6$ INFN Sezione di
Roma, P.le A. Moro 2, I-00185, Roma, Italy}
\affiliation
{$^7$ Universit\`a di Napoli Federico II, 
Dipartimento di Fisica ``Ettore Pancini'', piazzale Tecchio 80, 80126 Napoli, Italy}
\affiliation
{$^8$ Aix Marseille Universit\'e CNRS, CPT, UMR 7332, 13288
 Marseille, France \\
Universit\`e de Toulon, CNRS, CPT, UMR 7332, 83957 La Garde, France}
\affiliation
{$^9$ European Gravitational Observatory (EGO), 
I-56021 Cascina (Pi), Italy}

\begin{abstract}
In this paper we study the behavior of the Casimir energy of a ``multi-cavity'' across the transition 
from the metallic to the superconducting phase of the constituting plates. Our analysis is carried out in the framework of the ARCHIMEDES experiment, aiming at measuring the interaction of the electromagnetic vacuum energy  with 
a gravitational field. For this purpose it is foreseen to modulate the Casimir energy of a layered structure composing a multi-cavity coupled system by inducing a transition from the metallic to the superconducting phase.
This implies a thorough study of the behavior of the   cavity, where normal metallic layers are alternated with superconducting layers, across the transition.
 Our study finds that, because of the coupling between the cavities, mainly mediated by the transverse magnetic 
modes of the radiation field, the variation of energy across the transition can be very large.
\end{abstract}



\maketitle

\section{Introduction}
The ARCHIMEDES experiment \cite{erico_2014} is designed for testing whether the energy of vacuum fluctuations, 
foreseen by quantum electrodynamics, contributes to gravity, through the coupling demanded by quantum field theory 
in curved spacetime \cite{bimo_2006,fulling_2007,bimo_2008,espo_2008}, where the Einstein tensor is taken to 
be proportional to the expectation value of the regularized and renormalized energy-momentum tensor of matter 
fields. The idea is to weigh the vacuum energy stored in a rigid Casimir cavity made by parallel conducting plates, 
by modulating the reflectivity of the plates upon inducing a transition from the metallic to the superconducting phase 
\cite{erico_2014}. In order to enhance the effect, a multilayer cavity is considered, obtained by superimposing many cavities. 
This structure is natural in the case of crystals of type-II superconductors, particularly cuprates, being composed by 
Cu-O planes, that undergo the superconducting transition, separated by nonconducting planes. A crucial aspect 
to be tested is thus the behavior of the Casimir energy \cite{cas53} for a multi-cavity when the layers undergo 
the phase transition from the metallic to the superconducting phase. Until now only the case of a cavity having a single layer 
that undergoes the superconducting transition was considered, the other reflecting plate being just 
metallic (not superconducting), in Refs. \cite{bimonte051,bimonte052}.
The generalization to the case of a system of coupled  superconducting layered cavities is still lacking. With respect 
to the ARCHIMEDES project the main goal is to study the possibility of enhancing the modulation factor 
$\eta=\frac{\Delta E_{cas}}{E_{cas}}$ were $\Delta E_{cas}$ is the difference of Casimir energy in normal and superconducting 
states. The value obtained in Ref. \cite{bimonte051,bimonte052}, considering a cavity with a single superconducting layer and 
a transition temperature of about 1 K is $\eta_l\approx 10^{-8}$. This value was compliant with a previous experiment devoted 
to ascertain the vacuum energy contribution to the total condensation energy \cite{low-noise,allocca}, but it 
is not sufficient to prove the weight of the vacuum, because it is in absolute too small. It is therefore necessary to consider high-$T_c$ superconductors where condensation energy is much higher and also the absolute value of vacuum energy variation is expected to be correspondingly larger.
 
On the other hand, in Ref. \cite{kempf}, considering a cavity based on a high-$T_c$ layered superconductor, a factor as high as $\eta_h=4\cdot10^{-4}$ has been estimated, 
under the approximation of flat plasma sheets at zero temperature, no conduction in normal state (here $E_{cas}$ is the 
energy of the ideal cavity) and charge density of $n = 10^{14}$\,cm$^{-2}$. The ARCHIMEDES sensitivity 
is expected to be capable of ascertain the interaction of gravity and vacuum energy also for values lower than   
$\eta_h=4\cdot10^{-4}$, up to $1/100$ of this value \cite{erico_2014}. Clearly it is important to 
understand more firmly if dealing with layered superconducting structures the modulation depth can be sufficiently 
high. This is the study of the present paper. Considering in particular the multi-layer cavity, the general assumption 
adopted so far has been that the Casimir energy obtained by overlapping many cavities is the sum of the 
energies of each individual cavity. This is true if the {\em distances} between neighboring cavities 
are large (in the sense that the thickness of each metallic layer separating the various cavities is very large with respect 
to the penetration depth of the radiation field). Of course, this is no longer true if the thickness of these metallic 
inter-cavity layers gets thinner and thinner. The evaluation of the Casimir energy for such a configuration is the subject of the present study. It is worth  stressing that this is only a first step because in the final version 
ARCHIMEDES experiment will make use of high-$T_c$ superconducting oxides with a built-in layered structure, 
like YBa$_2$Cu$_3$O$_{7-x}$, for which a complete theory is as yet unavailable.

Having this in mind, we start with a thorough analysis of two and three coupled Casimir cavities, 
made by traditional BCS (low-$T_c$) superconducting material (niobium), so as to deal with relatively manageable 
and well established formulas. On trying to preserve a macroscopic approach, we limited our study to thicknesses 
between 10 and 100\,nm. In the following, referring to Fig.\,1, $d_i$ is the distance of the $i-th$ cavity from the 
$(i-1)-th$, (thickness of the $i-th$ cavity), within the slabs $1,3$ and $5$ there is vacuum while the zones $0,2,4$ and $6$ 
are made of niobium. The thicknesses of the zones $0$ and $6$ are assumed to be infinite. Although this choice is dictated 
by simplicity and by the well-established superconducting properties of Nb, this is a first necessary step
to prepare future studies aimed at considering the more elaborate case of high-$T_c$ superconductors, as required by the roadmap of the ARCHIMEDES experiment.

Section II studies the Casimir energy of a multilayer cavity, while Sec. III evaluates the Casimir 
energy in the normal and superconducting phases. Variation of the energy in the transition is obtained 
in Sec. IV, including a detailed numerical analysis of the Matsubara zero-mode contribution.
Section V extends this scheme to the three-layer configuration, and concluding remarks are made 
in Sec. VI, while relevant details are given in the Appendices.

\begin{center}
 \begin{figure}[ht]
     \centering
    \includegraphics[width = .65\textwidth]{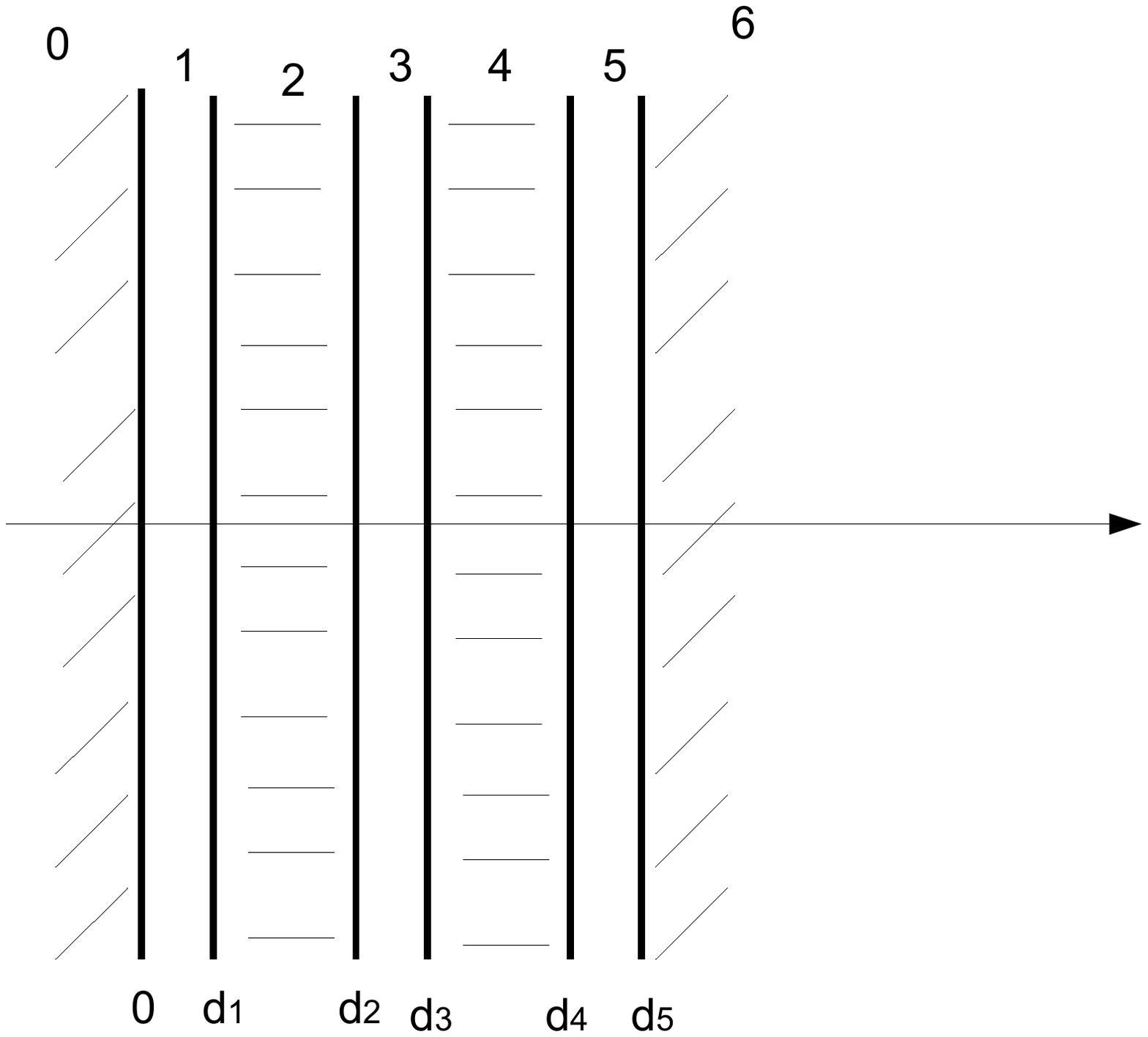}
   \caption{A three layer cavity. In the 0,2,4, and 6 zone there is Nb; in the 1,3,5 vacuum.  
$d_i$ is the thickness of the $i-th$ slab}

  \end{figure}
\end{center}

\section{The Casimir energy of a multilayer cavity}

As it is customary \cite{bimonte051,bimonte052}, at finite temperature, the Casimir variation across
the transition from a metallic to a superconducting phase is obtained as the difference between the free Casimir energy 
in the metallic state and the same after the transition to superconducting state takes place:
$\delta E(T)= {E_n}-{E_s}$. 
The energy per unit area of a single cavity, $(012)$ in Fig.\,1, can be written, at finite temperature $T$, 
as the sum of the contributions  of the transverse electric ($TE$) and 
transverse magnetic ($TM$) modes (see, for example, \cite{bord09}) :
\begin{equation} 
E[d_1,d_2]=\, {k_B
\,T}\sum_{l=0}^{~\infty\,~\prime}\int \frac{ d {\bf k_{\bot}}}{(2
\pi)^2} \,\left(\log{\Delta^{TE}(\xi_l)}+\log
{\Delta^{TM}(\xi_l)}\right)=:\sum_{l=0}^{\infty}  E[l,d_1,d_2]\; \label{fint1}
\end{equation}
where $\xi_l=2\pi l k_B T$ are the Matsubara frequencies, $k_B$ is the Boltzmann constant, 
$l=0,1,2,\ldots$, the superscript $^\prime$ on the sum means that the zero mode must be multiplied by a factor 
$\frac{1}{2}$,
\[
\Delta_{TM}(\xi_l) = \left(r_{TM}^{0,1}(\xi_l) r_{TM}^{1,2}(\xi_l) e^{-2 d_1 K_1}+1\right),   
\Delta_{TE}(\xi_l) = \left(r_{TE}^{0,1}(\xi_l) r_{TE}^{1,2}(\xi_l) e^{-2 d_1 K_1}+1\right)
\]
and the reflection coefficients $r_{(TM,TE)}^{i,j}(\xi_l)$ are given by (see \cite{bord09}): 
$$r_{TM}^{i,j}(\xi_l)=\frac{\epsilon_i(\xi_l)K_j(\xi_l)-\epsilon_j(\xi_l)K_i(\xi_l)}
{\epsilon_i(\xi_l)K_j(\xi_l)+\epsilon_j(\xi_l)K_i(\xi_l)}, ~r_{TE}^{i,j}(\xi_l)
=\frac{K_j(\xi_l)-K_i(\xi_l)}{K_j(\xi_l)+K_i(\xi_l)},
$$
with $K_i(\xi_l)=\sqrt{k_\perp^2+\epsilon_i(\xi_l) \xi^2_l}$. We point out that our approach captures the 
relevant length scale of a superconductor, the London penetration depth $\lambda_L$, through the expression of 
$\epsilon_i(\xi_l)$ in terms of the correction to the optical conductivity
when entering the superconducting state, $\delta\sigma_{BCS}(i\xi)$, see below and Appendix B. In particular, for
$\xi_l\to 0$, we have $\epsilon_i(\xi_l) \xi^2_l\to \lambda_L^{-2}$.

We characterize the properties of the $i-th$ material  trough the dielectric function 
$\epsilon_i(\xi_l)$ and  the change in the Casimir energy is given simply by the modification 
of the $\epsilon(\xi_l)$ due to the transition \cite{bimonte051,bimonte052}. 
As we said, in the following we report calculations for the case in which the material is Nb 
and the spacer is vacuum (the modifications introduced by a dielectric spacer
deserve a separate study).

To obtain the formulas for two and three cavities we solve the problem by imposing the 
continuity of the tangential component of the $\vec{E}$ and $\vec{H}$ fields (non-magnetic media) 
and the normal component of the $\vec{D}$ and $\vec{B}$ at the interface \cite{bord01,jack98}. 
Thus, for example, in the case of the three cavities (012-234-456) in Fig.\,1 we have that the 
$\Delta$ functions appearing in (\ref{fint1}) are the determinant of the  matrix of the coefficients 
$M_{ij}$  ({just to give an idea we report the expression for the $TM$- modes in appendix A}) 
from which it is possible to extract the case of one, two, and three cavities by taking 
$(i,j)=1\ldots4$, $(i,j)=1\ldots8$, $(i,j)=1\ldots12$, respectively:
$$
\Delta_{(TM,TE)}={\rm det}(M_{(TM,TE)}^{ij}).
$$
In the following we will omit the subscript $TM,TE$ if no ambiguity is generated. All the formulas 
for the two cases can be obtained using respectively the $TM$ or $TE$ reflection coefficients.
Defining (no summation over repeated indices)
$$
{E}^{ijl} =  r^{i,j} r^{j,l} e^{-2 d_j K_j}+1,~F^{ijl}  = r^{i,j}e^{-2 d_j K_j}+r^{j,l},  ~
G^{ijl} = r^{i,j} +e^{-2 d_j  K_j}r^{j,l},~  H^{ijl} =  e^{-2 d_j  K_j}+r^{i,j}r^{j,l} ,
$$
we have for the single cavity $(012)$ in Fig.\,1
\begin{equation}
\Delta^{(1)} = {E}^{012}; \label{eq:ene1}
\end{equation}
for two cavities $(012-234)$:
\beq
\Delta^{(2)} &=& {E}^{012}{E}^{234}+e^{-2(d_2k_2)}F^{012} G^{234}=:{E}^{012}{E}^{234}
+I^{(2)}_{012} \mbox{ ~~~   and} \label{eq:ene2}  \\ 
\log\Delta^{(2)} &=&\log\left({E}^{012}{E}^{234}\right) +\log\left(1+\frac{I^{(2)}_{012}}
{{E}^{012}{E}^{234}} \right), 
\label{eq:delta2}
\eeq
and for the three cavities:
\begin{eqnarray}
\Delta^{(3)} &=& E^{012}E^{234}E^{456}+
 e^{-2(d_2k_2+d_4k_4)}F^{012} H^{234}G^{456}   \nonumber \\
&& +e^{-2 d_2k_2}E^{456}F^{012} G^{234}+ 
e^{-2 d_4k_4}E^{012}F^{234} G^{456} \nonumber   \\
&&=:E^{012}E^{234}E^{456}+I^{(3)}+E^{456}I^{(2)}_{012}+E^{012}I^{(2)}_{234}, \label{eq:ene3} \\
\log\Delta^{(3)} &=&\log\left({E}^{012}{E}^{234}{E}^{456}\right) +
\log\left(1+\frac{I^{(3)}}{E^{012}E^{234}E^{456}} \right)  \nonumber \\
&&+ \log\left(1+\frac{E^{456}I^{(2)}_{012}+E^{012}I^{(2)}_{234}}
{E^{012}E^{234}E^{456}+I^{(3)}} \right).
\end{eqnarray}
In this way, when $d_2\rightarrow\infty$  [see Eq.\,(\ref{eq:ene2})] $I^{(2)}_{012}\rightarrow0$ and
$$
\log{ \Delta^{(2)}} = \log{{E}^{012}{E}^{234}}=\log{{E}^{012}}+\log{{E}^{234}}.
$$ 
That is to say, when the two cavities are far away their energy is simply the sum of the individual 
contributions. In this respect the second term on the right of Eq.\,(\ref{eq:ene2}), $I^{(2)}$,
can be seen as the energy due to the coupling of the two cavities $(012)-(234)$. 

When $d_1=d_3=d_5$, $d_2=d_4$, $\epsilon_0=\epsilon_2=\epsilon_4$, $\epsilon_1=\epsilon_3=\epsilon_5$ 
we obtain $ {E}_{TM}^{012}={E}^{234},~F^{012} =F^{234}$ so that we can omit the subscripts:

\begin{equation}
\log\Delta^{(2)} =\log\left(\Delta^{(1)} \right)^2 +\log\left(1+\frac{I^{(2)}}
{\left(\Delta^{(1)} \right)^2} \right). 
\label{eq:delta22}
\end{equation}

For the three cavities $(012-234-456)$, formulas are written so as to make evident the 
contribution to the energy resulting from the sum of the energies of the single cavity, with respect to 
the one coming from the coupling of the two possible pairs of cavities $(012-234),~(234-456)$, 
and the one coming from the coupling of the three: $I^{(3)}$. Thus, under the previous hypothesis,
\beq
\Delta^{(3)}  &=& \left(\Delta^{(1)} \right)^3+2I^{(2)}+I^{(3)}  \mbox{ and we can write } \nonumber \\
\log\Delta^{(3)}&=&\log \left(\Delta^{(1)} \right)^3+\log\left(1+2\frac{I^{(2)}}
{(\Delta^{(1)})^3+I^{(3)}} \right)+\log\left(1+\frac{I^{(3)}}{(\Delta^{(1)})^3} \right). 
\label{eq:delta3}
\eeq
In a sense, we are writing the energy as a sum of the energy of the single cavity plus the 
coupling energy between the nearest neighbor, plus the coupling energy among the second nearest neighbor 
and so on. In this way we will have a clear indication of the strength of the coupling between the cavities at 
the various orders.
{{  As far as we know this way of displaying the various contribution to the Casimir energy  has been obtained for the first time in \cite{Kenneth_2006} where the so called $TGTG$ formula ($T$ being the Lippmann-Schwinger $T$  operator  an $G$ the translation matrix), is used, see also \cite{Shajesh_2011, Teo_2012}. In our case it can be simply recovered by observing that the determinant of a $N\times N$ complex block  matrix can be obtained in terms of the determinants of its constituent blocks  \cite{Powell_2011}.}}

The Casimir energy in the superconducting phase is obtained by replacing, in the reflection 
coefficients, the expression of the dielectric function with the corresponding obtained using the 
BCS theory  \cite{zimmermann91,Bimonte:2009rb}, see Appendix B. In the following we will characterize the 
dielectric properties of the material by means of the Drude model (but see conclusions): 
\begin{eqnarray*}
\epsilon(i \xi) &=& 1+\frac{\sigma(i\xi)}{\xi}, ~~~~~~~~~~~{\mbox{  with}} \\
\sigma_k(i \xi) &=& \frac{\sigma_{0k}}{\gamma+ \xi }, {\mbox{~~~~~~~~~~~~~~~~ for conducting materials  and }} \\
\sigma_k(i \xi) &=& \frac{\sigma_{0k}}{\gamma+ \xi }+ \delta\sigma_{BCS}(i\xi),
{\mbox{ for superconducting materials }},
\end{eqnarray*}
where the expression of $\delta\sigma_{BCS}(i\xi)$ is given in Appendix B (see \cite{Bimonte:2009rb}).

Thus
\begin{eqnarray*} 
\delta E(T) &=&  E_n(T) -  {E_s}(T) \\
&=&\, \frac{k
\,T}{2}\sum_{l=0}^{~\infty\,~\prime}\int \frac{ d {\bf k_\bot}}{(2
\pi)^2} \,\left(\log \frac{\Delta^{(k)}_{n,TE}}{\Delta^{(k)}_{s,TE}}+\log
\frac{\Delta^{(k)}_{n,TM}}{\Delta^{(k)}_{s,TM}}\right)=:\sum_{l=0}^{~\infty}\delta E[l,d_1,d_2]. \label{fint2}
\end{eqnarray*}
where $\Delta^{(k)}_{n,TE,TM}$, $ \Delta^{(k)}_{s,TE,TM}$ are the generating functions
(this nomenclature denotes here just the determinant of the matrix whose zeros
provide, implicitly, the allowed energies) of 
the normal and superconducting phases, and, depending on how many cavities we are considering 
$(1,2,\mbox{ or }3)$ we must take  $k=1,2,\mbox{ or }3$ respectively.

\section{Casimir energy in the normal and superconducting phases}

All results described hereafter are obtained for Nb, and
we use the following values for the critical temperature and plasma frequency  
$T_c= 9.25$\,K, $\hbar \omega_{nio} = 9.268\,~eV$ and work at the temperature $T=9.157$\,K.
We start by choosing  $d_1=300$\,nm, and $d_2=600$\,nm, so as to have results that can be compared with standard 
formulas. 

We find for the energy in the normal phase ${E_n}$, for fixed $d_1, d_2$, and different 
values of number of Matsubara modes $(n_{mod})$: 
${E_n}^{n_{mod}}[ d_1, d_2] = \sum_{j=0}^{n_{mod}}  {E_n}[j,d_1,d_2]$ {{(N.B. in the following all the quoted numbers that concern energy or difference of energy are in $\frac{\mathrm J}{\mathrm m^2}$): }}

\bigskip

\centerline{
$
\begin{array}{|c|c|}
\hline
n_{mod}&
            {E_n}^{n_{mod}}[300,600]  \cdot 10^{8}~\\   \hline
 100 & -1.6577 \\   \hline
 200 & -2.2248 \\   \hline
 300 & -2.3760 \\   \hline
 400 & -2.4122 \\   \hline
 500 & -2.4203 \\   \hline
 1000 & -2.4226 \\   \hline  
\end{array} 
$}

\bigskip

Thus, at least $500$ Matsubara modes are necessary to obtain a result stable at the second decimal digit.

These results can be compared with the approximate result obtained by Bordag et al.  
\cite{bord09} ($t_0=\frac{c \,{\hbar}}{2 \, {a} \, {k_B T} },D_0=\frac{{c}}{{a} \, {\omega_p}}$):
\begin{eqnarray*}
E &=& \frac{ \ {c} \ {\hbar}}{\ 8 \pi {a}^3}
   \left[\ { D_0}^2 \left(\frac{3 \zeta (3)}{\ { t_0}}-\frac{4 \pi
   ^3}{25}\right)+ 
 \ { D_0} \left(\frac{2 \pi ^3}{45}-\frac{\zeta
   (3)}{\ { t_0}}\right)+\frac{\zeta (3)}{2 \ { t_0}}-\frac{\pi^3}{90}\right]\\
   &=& -1.258\cdot10^{-8}.
\end{eqnarray*}
As expected, because of the strong suppression of the exponential for large $d_2$, 
the contribution of the {\em coupling} term between the two cavities (012)-(123) is about ten 
orders of magnitude smaller than the energy obtained from each cavity. With 
\beq
E[d_1,d_2] &=& \frac{k\,T}{2}\sum_{l=0}^{~\infty\,~\prime}\int \frac{ d {\bf k_\bot}}{(2
\pi)^2}\left[ \log\left({E}^{012}{E}^{234}\right) +\log\left(1+\frac{I^{(2)}_{012}}
{{E}^{012}{E}^{234}} \right)\right]_{TM} +[{TE}]  
\\
&=:& E^{(2)}_{TM}[d_1,d_2] +C^{(2)}_{TM}[d_1,d_2] + E^{(2)}_{TE}[d_1,d_2] +C^{(2)}_{TE}[d_1,d_2] 
\\
&=:& E^{(2)}[d_1,d_2] +C^{(2)}[d_1,d_2] \nonumber
\label{eq:CC2}
\eeq
we obtain ($n_{mod}=500$):
$$
{E_n}^{(2)} [300,600]=-2.42035\cdot10^{-8}    ;~~~{C_n}^{(2)} [300,600] = -2.30743\cdot10^{-17} .
$$

Indeed, having $d_1\ll d_2$, the total energy is simply the sum of the energies of the two cavities: 
$$\frac{ {E_n}^{(2)} [300,600] }{2}  =\frac{-2.42035}{2}\cdot10^{-8} =-1.2102\cdot10^{-8} \approx -1.258\cdot10^{-8} $$

\bigskip

In the superconducting phase we have, more or less, the same behavior:
\bigskip

\centerline{
$
\begin{array}{|c|c|}
\hline
n_{mod}&
            {E_{s}}^{n_{mod}}[300,600]  \cdot 10^{8}\\   \hline
 100 & -1.6584 \\   \hline
 200 & -2.2255 \\   \hline
 300 & -2.3767 \\   \hline
 400 & -2.4129 \\   \hline
 500 & -2.4210 \\   \hline
 1000 & -2.4232 \\   \hline  
\end{array} 
$}
\bigskip
The Casimir energy is always greater than the corresponding energy in the normal phase so that, 
as expected, the difference is always positive. 
Once more, the contribution from the energy of the two cavities is much larger than the contribution 
from the coupling, but ``only'' by about four orders of magnitude:
$$
{E_s}^{500}[300, 600] =-2.4210\cdot10^{-8},
{E_s}^{(2)}[ 300, 600]= -2.4208\cdot10^{-8},  {{C_s}}^{(2)}[ 300, 600]=-2.0979\cdot10^{-12}.
$$

\section{Variation of the energy across the transition}

In computing the difference in energy between the two phases,
we find that a few tens $(50)$ of modes are sufficient to obtain good values. This is a consequence
of the fact that the high-energy part of the spectrum is essentially the same in the metal and in the superconductor,
making the energy difference a quantity that converges much more rapidly than the individual terms, as a function of the 
upper cutoff in the Matsubara frequency, $n_{mod}$.

On defining $\delta E^{(2)}+{\delta C^{(2)}}$, i.e. 
$$
\delta E^{n_{mod}} [d_1,d_2]= {E_n}^{(2)}[d_1,d_2]-{E_s}^{(2)} [d_1,d_2]  +
{C_n}^{(2)}-{C_s}^{(2)}=:\delta E^{(2)}+{\delta C^{(2)}}
$$
as the difference between the terms coming from the energy of the two cavities 
in the normal and superconducting phase, plus 
the difference between the values of the coupling in the two phases respectively, we have:
  
\bigskip

\centerline{
$
\begin{array}{|c|c|c|c|}
\hline
             n_{mod}&\de^{n_{mod}} [300,600] \cdot 10^{12}&\delta{E^{(2)}}\cdot 10^{12}& \delta{C^{(2)}} \cdot 10^{12} \\   \hline
 10 &6.54447& 4.44655 &2.09792 \\   \hline
 30 & 6.55295& 4.45504 & 2.09792 \\   \hline
 50 &  6.55406 &4.45614 & 2.09792\\   \hline
 100 & 6.55461&4.45669 & 2.09792 \\   \hline  
\end{array} 
$}

\bigskip
We note that $\delta E^{(2)}$ is of the same order of magnitude of $\delta C^{(2)}$, 
and when $d_1=d_2$~~ $\delta C^{(2)}$ is about two orders of magnitude larger:
\beq
\de^{50}[ 50,50]  &=& 1.11259\cdot 10^{-9} =  1.24592\cdot 10^{-11} +  1.10013\cdot 10^{-9} \\
\de^{50}[10,10] &=& 9.3812\cdot10^{-9}= 2.34485\cdot10^{-11}+9.35775\cdot10^{-9}.
\eeq

\subsection{The Matsubara zero-mode contribution}

It turns out that this unexpected behavior is due to the contribution from the Matsubara zero mode.
This is evident in the following table where we report, for the $n-th$ Matsubara-mode, 
the values of the Casimir energy in the normal and superconducting phase, and their difference ($d_1=d_2=100$\,nm):
\begin{center}
{\footnotesize{
$$
\begin{array}{|c|c|c|c|c|c|}  \hline
n& E[n,100,100] =& \{E^{(2)}+C^{(2)}\}_{TE}&  \{E^{(2)}+C^{(2)}\}_{TM}  & \{E^{(2)}+C^{(2)}\}_{TE+TM}\\ \hline

& E_n & 0+0 & -6.039\cdot10^{-10} -2.175\cdot10^{-13} &  -6.041\cdot10^{-10} \\ \cline{2-5}

\multirow{2}{*}{0} & E_s& -6.284\cdot10^{-12} -1.647\cdot10^{-13} &  -6.039\cdot10^{-10}-2.692 
\cdot10^{-10} & -8.795\cdot10^{-10} \\ 
\cline{2-5}

& \de& 6.284\cdot10^{-12}+ 1.647\cdot10^{-13} & 0.0+ 2.690\cdot10^{-10} & 2.754\cdot10^{-10} \\ \hline

& E_n& -2.076\cdot10^{-10} -6.060\cdot10^{-12}  
& -1.207\cdot10^{-9} -1.235\cdot10^{-10}  & -1.545\cdot10^{-9} \\ \cline{2-5}
\multirow{2}{*}{1}& E_s& -2.087\cdot10^{-10} -6.048\cdot10^{-12}  
& -1.207\cdot10^{-9} -1.223\cdot10^{-10}  & -1.545\cdot10^{-9} \\\cline{2-5}

& \de& 1.161\cdot10^{-12} -1.228\cdot10^{-14}  & 9.480\cdot10^{-16}-1.235\cdot10^{-12}  & -8.536\cdot10^{-14}  \\ \hline

& E_n& -4.64934\cdot10^{-10} -8.59787\cdot10^{-13} & -1.19878\cdot10^{-9}, 
-5.58978\cdot10^{-12} &  -1.67017\cdot10^{-9} \\ \cline{2-5}

\multirow{2}{*}{10} & E_s&-4.64952\cdot10^{-10} -8.5958\cdot10^{-13} & -1.19878
\cdot10^{-9}-5.58814\cdot10^{-12}&-1.67018\cdot10^{-9} \\ 
\cline{2-5}

& \de&1.80615\cdot10^{-14} -2.07135\cdot10^{-16}& 2.31737\cdot10^{-16} -1.64377
\cdot10^{-15}& 1.64423\cdot10^{-14} \\ \hline

& E_n&-5.11734\cdot10^{-10} -1.87681\cdot10^{-13}& -1.1032\cdot10^{-9} -9.34075
\cdot10^{-13}&  -1.61606\cdot10^{-9} \\ \cline{2-5}

\multirow{2}{*}{50} & E_s &-5.11735\cdot10^{-10}-1.8768\cdot10^{-13} & -1.1032
\cdot10^{-9}-9.34068\cdot10^{-13}&-1.61606\cdot10^{-9} \\ 
\cline{2-5}

& \de&3.31224\cdot10^{-16} -1.22071\cdot10^{-18}&3.90994\cdot10^{-17}-6.84111
\cdot10^{-18}& 3.62262\cdot10^{-16} \\ \hline

\end{array}
$$}}
 \captionof{table}{Contributions of the $TE$ and $TM$ modes for different values of $n$}
\end{center}
and summing the first $50$ modes:
$$
\de^{50}[100, 100] =2.76004\cdot10^{-10} = 8.54145\cdot10^{-12} + 2.67463\cdot10^{-10}.
$$
A close look at the table makes it evident that the result is almost completely 
due to the contribution of the coupling term of the zero mode. 
Indeed, $C_{s,TM}^{(2)}$ is about 3 orders of magnitude larger than the corresponding in 
the normal case  $C_{n,TM}^{(2)}$ while all the other terms are of the same order of magnitude 
(in some case egual) so that in the difference they cancel each other.

\section{Energy of the three-layer configuration}

The behavior discussed in the previous section is confirmed for the three-layer configuration:
\beq
E[d_1,d_2,d_4] &=& \frac{k_B\,T}{2}\sum_{l=0}^{~\infty\,~\prime}\int \frac{ d {\bf k_\bot}}{(2
\pi)^2} \left[
\log\left({E}^{012}{E}^{234}{E}^{456}\right) +
\log\left(1+\frac{I^{(3)}}{E^{012}E^{234}E^{456}} \right)  \right. \nonumber \\
&& + \left. \log\left(1+\frac{E^{456}I^{(2)}_{012}+E^{012}I^{(2)}_{234}}{E^{012}E^{234}E^{456}+I^{(3)}} \right)
\right] _{TM}+[TE], \\
&=:& 
E^{(3)}[d_1,d_2,d_4] +C^{(3)}[d_1,d_2,d_4] +C^{(2)}[d_1,d_2,d_4]\nonumber
\label{eq:CC22} \\
E^{n_{mod}}[ d_1, d_2,d_4] &:=& \sum_{l=0}^{n_{mod}}  E[l,d_1,d_2,d_4], \nonumber
\eeq
To have a comparison between the formulae for two and three cavities let us compute the Casimir energy 
for the three layer when $d_4$ is very large. In this case, since the third cavity is distant from the other two, it decouples 
and the result would be the sum of the energy of a double cavity plus the energy of a third one. Indeed we find:
$$
\begin{array}{|c|c|c|c|c|}  \hline
E_{n }^{500}(100,100,300)  & E_{n }^{(3)} &C_{n }^{(2)} & C_{n }^{(3)} \\ \hline
-6.2636\cdot10^{-7} & -6.2574\cdot10^{-7} & -6.2390\cdot10^{-10} & 6.73512\cdot10^{-15}\\ \hline
\end{array}
$$
this is exactly three halves the energy of a double cavity:
\begin{equation}
E_{n }^{500}( 100, 100) =-4.17785\cdot10^{-7} \approx -\frac{2}{3}6.26365\cdot10^{-7}
=4.17577\cdot10^{-7}, 
\end{equation}
as expected. Of course, this is a consequence of the strong exponential suppression present 
in this term, see the expression of $I^{(3)}$ in Eq.(\ref{eq:ene3}).
Taking $d_4=d_1=d_2=100$\,nm we find
$$
\begin{array}{|c|c|c|c|c|}  \hline
E_{n }^{500}(100,100,100)  & E_{n }^{(3)} &C_{n }^{(2)} & C_{n }^{(3)} \\ \hline
-6.26989\cdot10^{-7} & -6.25742\cdot10^{-7} & -1.25193\cdot10^{-9} & 4.39881\cdot10^{-12}\\ \hline
\end{array}
$$
thus the contribution due to the coupling of the three cavities is three orders of magnitude larger 
than in the previous case but still, for the normal Casimir energy, very much smaller than the sum of 
the energies of the three individual cavities.

Once again, things are different when computing the difference between the energy in the normal 
and superconducting phase. Indeed, in this case the contribution from the $n-th$ mode is, 
with obvious significance for the indicated symbols: 
$$
\begin{array}{|c|c|c|c|c|}  \hline
n & \de(n,100,100,100) & \de^{(3)}& \delta C^{(2)}  & \delta C^{(3)} \\ \hline
0 &5.47751\cdot 10^{-10} &  9.42558\cdot 10^{-12} &  5.84966\cdot 10^{-10 } &  -4.66401\cdot 10^{-11} \\ \hline 
1 &-7.53124\cdot 10^{-13} &  1.74333\cdot 10^{-12} &  -2.57231\cdot 10^{-12} &  7.58578\cdot 10^{-14}\\ \hline 
10 &2.3738\cdot 10^{-14} &  2.7440\cdot 10^{-14} &  -3.7044\cdot 10^{-15} &  2.49605\cdot 10^{-18}\\ \hline 
100 &8.97745\cdot 10^{-17} &  9.15143\cdot 10^{-17} &  -1.73989\cdot 10^{-18} &  9.25498\cdot 10^{-23}\\ \hline 
\end{array}
$$

\bigskip
We immediately realize that even in this case the energy is due almost completely to the coupling 
of nearest cavities $\delta C^{(2)} $.  Note that the $\delta C^{(3)}$ term is about one order of 
magnitude smaller than the corresponding $\delta C^{(2)}$.
Summing on the first $n$ modes we find 

\bigskip

$$
\begin{array}{|c|c|c|c|c|}  \hline
n_{mod} & \de^{n_{mod}}(100,100,100)  & \de^{(3)}& \delta C^{(2)}  & \delta C^{(3)} \\ \hline
10 &5.47634\cdot10^{-10} & 1.2644\cdot10^{-11} & 5.81547\cdot10^{-10} & -4.6557\cdot10^{-11} \\ \hline 
50 &5.47789\cdot10^{-10} & 1.28122\cdot10^{-11} & 5.81533\cdot10^{-10 } & -4.6557\cdot10^{-11} \\ \hline 
100 &5.47800\cdot10^{-10} & 1.28236\cdot10^{-11} & 5.81533\cdot10^{-10} & -4.6557\cdot10^{-11} \\ \hline 
\end{array}
$$
For layers $10$\,nm thick we find:
$$
\begin{array}{|c|c|c|c|c|}  \hline
E_{n }^{500}(10,10,10)  & E_{n }^{(3)} &C_{n }^{(2)} & C_{n }^{(3)} \\ \hline
-1.134\cdot10^{-4} & -7.880\cdot10^{-5} & -3.641\cdot10^{-5} & 1.761\cdot10^{-6}\\ \hline
\end{array}
$$
and
$$
\begin{array}{|c|c|c|c|c|}  \hline
\de^{100}(10,10,10)  & \delta E^{(3)} &\delta C^{(2)} & \delta C^{(3)} \\ \hline
1.875\cdot10^{-8} & 3.517\cdot10^{-11} &2.166\cdot10^{-8} & -2.946\cdot10^{-9}\\ \hline
\end{array}
$$
{{
\begin{center}
 \begin{figure}[ht]
     \centering
    \includegraphics[width = .7\textwidth]{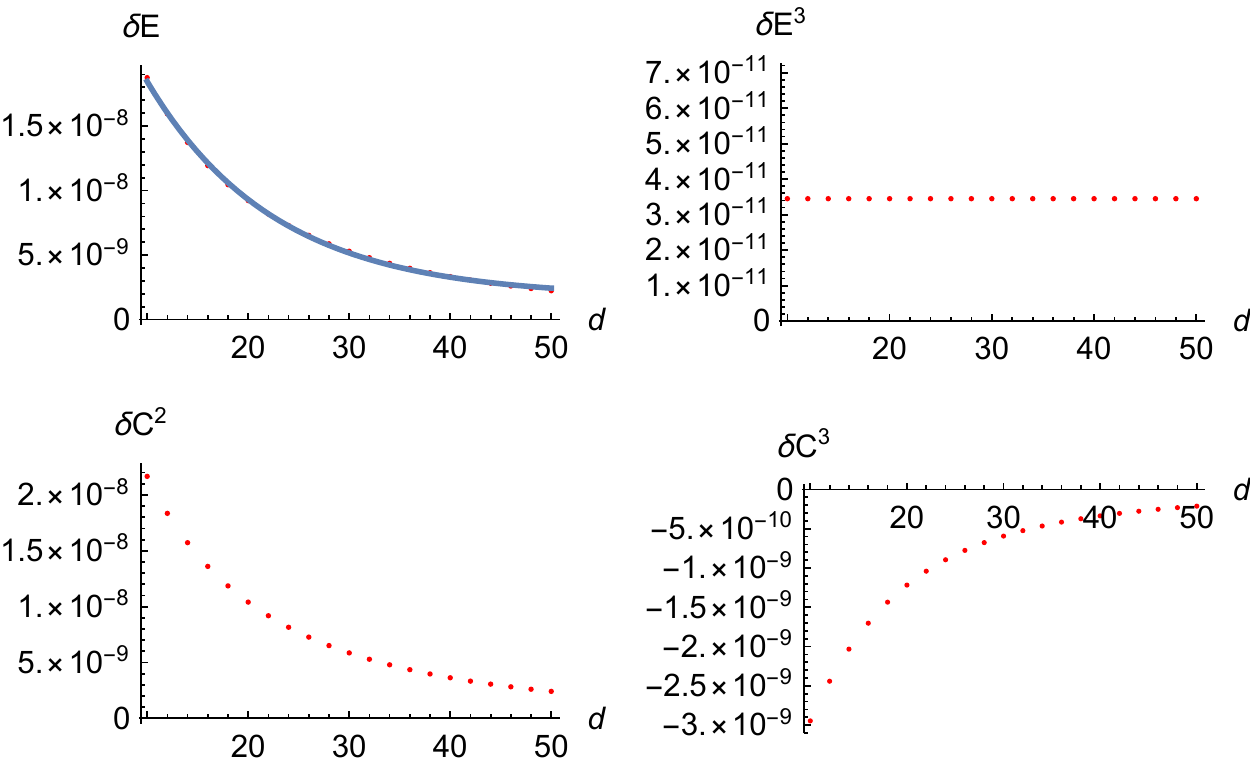}
   \caption{The behavior with respect to $d_2=d_4=d\in[10,50]~nm$ of the Casimir energy $\delta E$ and of the various components $\delta E^3,~\delta C^2$, and $\delta C^3$  for the three-layer configuration with $d_1=10~nm$. In the plot of $\delta E$ it is shown, also, the fitting curve.}
  \end{figure}
\end{center}
To give an idea of the dependence of the Casimir energy on the parameters $d_1,d_2,d_4$, we show in the figure 2 the contribution of the three terms $\delta E^3,~\delta C^2$, and $\delta C^3$ to the energy difference between the normal and the superconducting phase, $\delta E$, with respect to $d_2=d_4=d\in[10,50]~nm$ with  $d_1=10~nm$. In blue it is shown a fit of $\delta E$ obtained by means of the function $\delta E=a+b~ e^{-\left(\frac{x}{x_0}\right)}$
with $a=1.73\cdot 10^{-9}~J/m^2,~b=3.69\cdot 10^{-8}~J/m^2,~x_0=12.63~nm$. Note that the red dots are completely covered by the fitting curve. The only term that substantially depends on $d_1$ is $\delta E^3$, i.e. the sum of the energies of the single cavity whose thickness is $d_1$. On the contrary the other terms almost  depend on $d_2,d_4$ exclusively. Being $\delta E^3$ very much smaller than $\delta C^2$ and $\delta C^3$, this fit is very stable with respect the variation of $d_1$ see FIG. 3 where the same fitting curve is overimposed on the data relative to $d_1=500~nm$
\begin{center}
 \begin{figure}[ht]
     \centering
    \includegraphics[width = .7\textwidth]{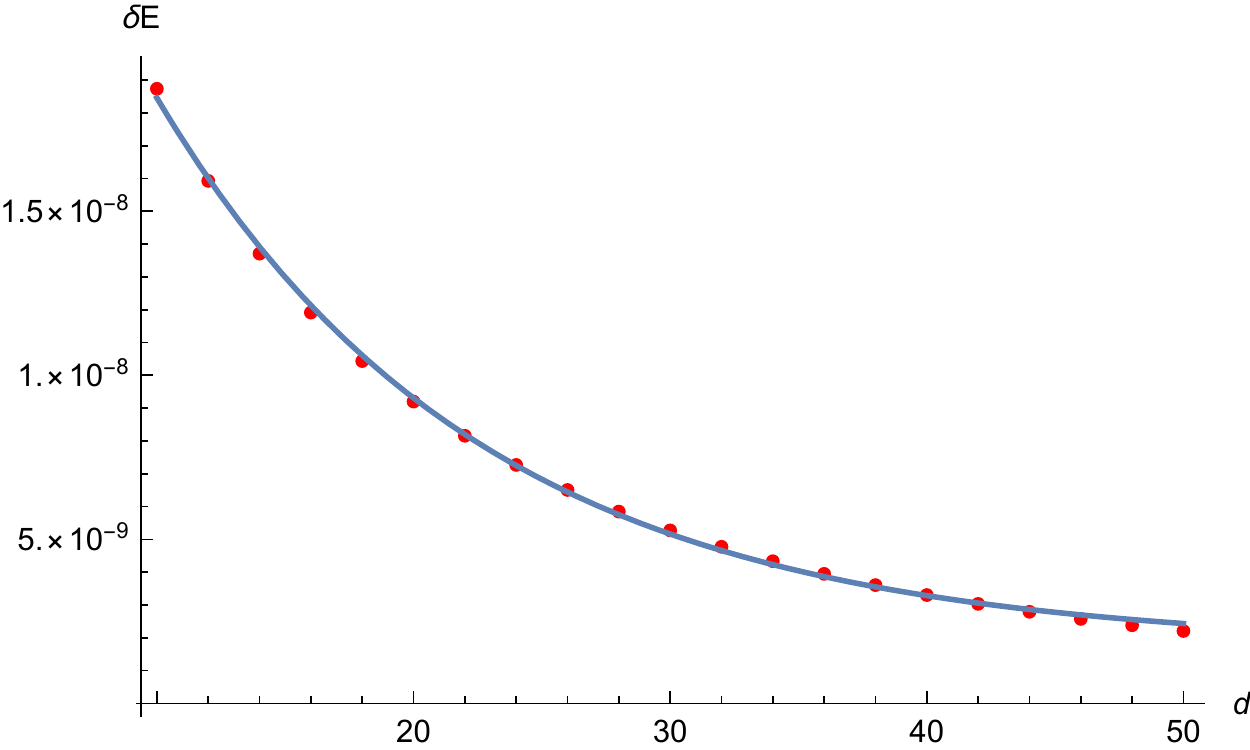}
   \caption{The behavior with respect to $d_2=d_4=d\in[10,50]~nm$ of the Casimir energy $\delta E$ for the three-layer configuration with $d_1=500~nm$ and the fitting curve with the parameters obtained for the case $d_1=10~nm$.}
  \end{figure}
\end{center}
}}

{{We conclude that, the contribution from  the coupling of the three cavities  being so large: $\delta C^3$ can turn out to be  only one order of magnitude smaller than $\delta C^2$, it will be therefore necessary to analyze the situation of four coupled cavities.}}

Some comments about the contribution of the $TM$ zero mode are in order at this point (in the following we 
will analyze the configuration of two coupled cavities but the generalization to three is straightforward). 
%
%
In the $\xi\mapsto0$ limit,  assuming vacuum between
the two superconducting layers (see appendix B), one obtains $r_{TM}^{i,j}=1$ for both the normal and superconducting case, so that: 
\begin{eqnarray*}
E^{(2)}_{TM,n} &=& E^{(2)}_{TM,s}=\frac{k_B\,T}{4}\int \frac{ d {\bf k_\bot}}{(2
\pi)^2} \log\left(  1-e^{-2d_1K_1}\right) \\
C^{(2)}_{TM,n} &=&\frac{k_B\,T}{4}\int \frac{ d {\bf k_\bot}}{(2\pi)^2} \log\left(  1-e^{-2d_2K_{0,n}}\right) \\
C^{(2)}_{TM,s} &=&\frac{k_B\,T}{4}\int \frac{ d {\bf k_\bot}}{(2
\pi)^2} \log\left(  1-e^{-2d_2K_{0,s}}\right) 
\end{eqnarray*}
where $d_i$ are measured in nm, $K_i$ in nm$^{-1}$, and 
$K_{i,n/s}=\sqrt{k^2_\perp+\alpha^2_{i,n/s}}$. We immediately realize that the contribution of the energies of the two cavities 
$E^{(2)}_{TM}$ is exactly the same in the normal and in the superconducting phase so that they cancel 
in the difference. On the contrary the contribution of the interaction terms $C^{(2)}_{TM}$ in the two phases are different 
thanks to the presence of $K_{0,n}$ and $K_{0,s}$ respectively. (Note that the dependence of these two terms on
$d_1$ cancels). Naturally, the strong suppression due to the exponential ensures that the main contribution to the integral 
comes from small wave numbers $k_\perp$ (hereafter measured in nm$^{-1}$),
taking $d_1=d_2=100$\,nm and 
$\alpha^2_1:=\sqrt{k^2_\perp+\left(\frac{\omega_{nio}}{ c}\right)^2}=\sqrt{k^2_\perp+0.00221}$\,nm$^{-1}$, 
$\alpha^2_{0,n}:=k^2_\perp$, and $\alpha^2_{0,s}:=\sqrt{k^2_\perp+\left(\frac{\omega_{s}}{ c}\right)^2}
=\sqrt{k^2_\perp+4.306\cdot 10^{-6}}$\,nm$^{-1}$, we recover the numbers in Table I. Of course the huge difference
reduce drastically for smaller values of $d_2$ because smaller values of $d_2$ allow for the contribution to the integral 
from larger values of $k_\perp$ so that the dependence on $\alpha^2_{i,n/s}$ is less evident. For example with 
$d_2=10$\,nm
we get $C^{(2)}_{TM,n} =1.78\cdot10^{-8}$\,J and $C^{(2)}_{TM,s} =-3.011\cdot10^{-8}$\,J. 
However in this case, even though the two terms are very much closer, their absolute value is larger 
so that they still give a strong contribution to the energy, thus the values of $\delta E$ is large, see Eqs.
(12) and (13). We expect that this behavior could change when a dielectric is inserted between the two layers.
%
%

\bigskip

\section{Concluding remarks}

In this paper we performed a series of numerical calculations aimed at the computation of the Casimir 
energy in the normal and superconducting phase for a multilayered cavity.
This is of particular interest for the ARCHIMEDES experiment aimed at weighing the vacuum energy of a 
multi-cavity by modulating the reflectivity of the constituting plates
from the metallic to the superconducting phase. As pointed out in \cite{erico_2014} 
with a single cavity and with a standard BCS superconductor  a ratio  $\eta=\frac{\Delta E_{cas}}{E_{cas}}\sim 10^{-8}$ is expected. For this value there would be no possibility for the experiment to detect 
the signal. However, and quite surprisingly, our results are orders of magnitude larger: 
We obtained a very large contribution from a term resulting from the coupling of nearest 
neighbor cavities in the superconducting phase. This strong enhancement of $\eta$
results from the use of a superconducting multi-layer (at least two) structure and it can be attributed to the strong contribution 
of the $TM$ Matsubara zero mode. From the point of view of the experiment these results are quite promising. 

The important role played by the static TM physically arises 
because, while a static electric field in a superconductor (and in a metal as well) is rapidly screened on short 
length-scales, the magnetic field parallel to the vacuum-Nb interface can penetrate over a substantial distance, 
set by the London penetration depth. This length is shortest in clean Nb, but is still of the order 
of tens of nm, and increases in the presence of impurities. It is not surprising therefore, that the zero-frequency 
TM mode links the various adjacent cavities, providing a substantial inter-cavity contribution to the Casimir energy.
Therefore, in computing the Casimir energy of a large number of overlapping cavities, it is 
necessary to take into account the contribution from the coupling of pairs of cavities 
that can lead to a strong enhancement of the effect. This behavior is confirmed in the case of a three-layer 
configuration where, in addition, the contribution of the coupling of the three 
cavities turns out to be about one order of magnitude smaller. At this stage we plan to obtain in a future work an 
estimate of the contribution of (at least) four coupled cavities. Because of the strong contribution of 
the zero mode we expect to be able to discriminate between the Drude or plasma model in computing the 
zero-mode contribution for the Casimir energy.
%
We wish to point out that, even though these results are 
encouraging, the shift in energy is still not as large as needed. Indeed  (see \cite{erico_2014}, 
and Refs. therein), to extract the signal we need an energy shift of the order of few joules. With this 
kind of configuration, even using a very thin layer, of the order of few nanometers, the energy shift is 
relatively small: $\de^{100}(1,1,1)=6.371\cdot10^{-8}$
This is a consequence of the smooth dependence of $\de$ on $d_2$. Indeed, for 
$d_2\leq10$\,nm, it can be fitted as (see \cite{bimonte052}):
$$\de^{100}(10,d_2,d_2)=\frac{\de_0}{1+\left(\frac{d_2}{D}\right)^s}$$
with $\de_0=9.29\cdot10^{-8},s=0.92, D=2.30~nm$. Thus, in this way we can gain at most one order 
of magnitude. This result strongly support our idea of obtaining such an improvement by using high-temperature 
superconducting oxides, like YBa$_2$Cu$_3$O$_{7-x}$. In this case, in fact, larger areas can be 
used (two orders of magnitude), a larger number of layer, $\sim 10^6$, can be assembled 
together, relying on the fine built-in layered structure of cuprates, with thickness 
of the order of 1\,nm. It is possible to work at high temperature, $\sim 100$\,K (gaining here a factor ten), 
and, possibly, other two order of magnitude can be gained from $\Delta T$. Of course, this prevision can 
prove to be too optimistic and for this reason the extension of the present analysis to such a situation 
is underway.

%


\begin{acknowledgments}
G.E. and C.S. are grateful to the Department of Physics ``Ettore Pancini'' 
of Federico II University, Naples, for hospitality and support.
\end{acknowledgments}

\begin{appendix}

\section{ }
For the case of the $TM$-modes the matching conditions give the following 
$12 \times 12$ matrix of coefficients:

\begin{center}
\scalebox{.65}{
$
M=\left(
\begin{array}{cccccccccccc}
 -\epsilon_0 & \epsilon_1 & \epsilon_1 & 0 & 0 & 0 & 0 & 0 & 0 & 0 & 0 & 0  \\
 -K_0 & K_1 & -K_1 & 0 & 0 & 0 & 0 & 0 & 0 & 0 & 0 & 0 \\
 0 & e^{d_1 K_1} \epsilon_1 & e^{-d_1 K_1} \epsilon_1 & -e^{-d_1 K_2}
   \epsilon_2 & -e^{d_1 K_2} \epsilon_2 & 0 & 0 & 0 & 0 & 0 & 0 & 0 \\
 0 & e^{d_1 K_1} K_1 & -e^{-d_1 K_1} K_1 & e^{-d_1 K_2} K_2 & -e^{d_1
   K_2} K_2 & 0 & 0 & 0 & 0 & 0 & 0 & 0 \\
 0 & 0 & 0 & e^{-K_2 x_2} \epsilon_2 & e^{K_2 x_2} \epsilon_2 & -e^{K_3
   x_2} \epsilon_3 & -e^{-K_3 x_2} \epsilon_3 & 0 & 0 & 0 & 0 & 0 \\
 0 & 0 & 0 & -e^{-K_2 x_2} K_2 & e^{K_2 x_2} K_2 & -e^{K_3 x_2} K_3 &
   e^{-K_3 x_2} K_3 & 0 & 0 & 0 & 0 & 0 \\
 0 & 0 & 0 & 0 & 0 & e^{K_3 x_3} \epsilon_3 & e^{-K_3 x_3} \epsilon_3 &
   -e^{-K_4 x_3} \epsilon_4 & -e^{K_4 x_3} \epsilon_4 & 0 & 0 & 0 \\
 0 & 0 & 0 & 0 & 0 & e^{K_3 x_3} K_3 & -e^{-K_3 x_3} K_3 & e^{-K_4 x_3}
   K_4 & -e^{K_4 x_3} K_4 & 0 & 0 & 0 \\
 0 & 0 & 0 & 0 & 0 & 0 & 0 & e^{-K_4 x_4} \epsilon_4 & e^{K_4 x_4}
   \epsilon_4 & -e^{K_5 x_4} \epsilon_5 & -e^{-K_5 x_4} \epsilon_5 & 0  \\
 0 & 0 & 0 & 0 & 0 & 0 & 0 & -e^{-K_4 x_4} K_4 & e^{K_4 x_4} K_4 &
   -e^{K_5 x_4} K_5 & e^{-K_5 x_4} K_5 & 0 \\
 0 & 0 & 0 & 0 & 0 & 0 & 0 & 0 & 0 & e^{K_5 x_5} \epsilon_5 & e^{-K_5 x_5}
   \epsilon_5 & -e^{-K_6 x_5} \epsilon_6 \\
 0 & 0 & 0 & 0 & 0 & 0 & 0 & 0 & 0 & e^{K_5 x_5} K_5 & -e^{-K_5 x_5} K_5 &
   e^{-K_6 x_5} K_6 \\
\end{array}
\right).
$
}
\end{center}
\bigskip
Computing the determinant of the minors of dimensions $4,8$, and $12$ respectively we 
obtain Eqs.\,(\ref{eq:ene1},\ref{eq:ene2},\ref{eq:ene3}).

\section{ }
On writing
\begin{equation*}
\epsilon(i \xi) = 1+\frac{\sigma(i\xi)}{\xi}
\end{equation*}
where $\sigma(i\xi)$ is the conductivity along the imaginary frequencies, we will obtain the dielectric 
function in the Drude model for the normal case simply by taking
\begin{equation*}
\sigma(i \xi) = \frac{\omega^2_{p}/4\pi}{\gamma+ \xi }
\end{equation*}
with $\omega^2_p=4\pi n e^2/m$ the plasma frequency and $\gamma$ the relaxation parameter.
While in the superconducting phase the conductivity can be written as \cite{Bimonte:2009rb} 
\begin{equation*}
\sigma(i\xi) = \frac{\omega_{p}}{\gamma+ \xi }+\delta\sigma_{BCS}(i\xi)
\end{equation*}
where the correction within the $BCS$ model is given by {{(in the following $\hbar=1$)}}
\begin{eqnarray*}
\delta\sigma_{BCS}(i\xi) &=& \frac{\sigma_0\gamma}{\xi}\int_{-\infty}^{+\infty}{\tanh{\left(\frac{E}{2T}\right)}
Re[G_+(i\xi,\eta)]\frac{d\eta}{E}}, \\ \nonumber \\
G_+(z,\eta) &=&\frac{\eta^2Q_+(z,E)+A_+(z,E)(Q_+(z,E)+i\gamma)}{Q_+(z,E)[\eta^2-(Q_+(z,E)+i\gamma)^2} , \\ \nonumber \\
A_+(z,E)&=&E(E+z)+\Delta^{2} , \\
Q_+^2(z,E)&=&(E+z)^2-\Delta^{2},\\
E&=&\sqrt{\eta^2+\Delta^2}.
\end{eqnarray*}
To obtain the reflection coefficients for the zero mode we have to compute 
the limit $\xi\rightarrow0$. In this way, considering that in the $1, 3$, and $5$ 
regions there is vacuum, we find
$$
\lim_{\xi\rightarrow  0}r_{TM}^{i,j}(i\xi)= 1,~~\lim_{\xi\rightarrow  0}r_{TE}^{i,j}(i\xi)=0
$$
in the conducting phase. Noting that, when $\xi\rightarrow0$,
$\delta\sigma_{BCS}(z)$ can be approximated by \cite{Bimonte:2009rb} 
$$
\delta\sigma_{BCS}(i\xi)\approx\omega^2_s/\xi ,
$$ 
with
\begin{eqnarray*}
\omega_s^2=\frac{\omega^2_p}{\gamma}\left( \pi \Delta \tanh{\frac{\Delta}{2 k_b T}}-4\gamma\Delta^2
\int_0^\infty{
\frac{ \tanh{\frac{\sqrt{\Delta^2+x^2}}{2 k_b T} }}{\sqrt{\Delta^2+x^2}(\gamma^2+x^2)} dx}\right),
\end{eqnarray*}
we obtain for $r_{TM,TE}^{i,j}$ in the superconducting phase:
$$
\lim_{\xi\rightarrow  0}r_{TM}^{i,j}(i\xi)= 1,~~\lim_{\xi\rightarrow  0}r_{TE}^{i,j}(i\xi)
=\frac{k_\perp-\sqrt{k_\perp^2+\omega_{s i}^2} }{k_\perp+\sqrt{k_\perp^2+\omega_{s i}^2}} .
$$
\end{appendix}

\end{document}